\documentstyle[11pt,IAUS212,twoside,epsf]{article}

\def\edcomment#1{\iffalse\marginpar{\raggedright\sl#1\/}\else\relax\fi}
\reversemarginpar

\begin{document}
\vspace*{1cm}
\title{Evolution of zero-metallicity massive  stars}
 \author{Paola Marigo}
\affil{Dipartimento di Astronomia, Universit\`a di Padova, Vicolo dell'Osservatorio 2, I-35122, Padova, Italia}
\author{Cesare Chiosi}
\affil{Dipartimento di Astronomia, Universit\`a di Padova, Vicolo dell'Osservatorio 2, I-35122, Padova, Italia}
\author{L\'eo Girardi}
\affil{Osservatorio Astronomico di Trieste, Via Tiepolo 11, I-34131, Trieste, 
Italia}
\author{Rolf-P. Kudritzki}
\affil{Institute for Astronomy, University of Hawaii, 2680 Woodlawn Drive, Honolulu, HI 96822}

\begin{abstract}
We discuss the evolutionary properties of
primordial massive and very massive stars, supposed to have formed from
metal-free gas. Stellar models are presented over a large range of
initial masses ($8 M_{\odot} \la M_{\rm i} \la 1000  M_{\odot}$),
covering the hydrogen- and helium-burning phases up to the 
onset of carbon burning. In most cases the
evolution is followed at constant mass. 
To estimate the possible effect of mass loss via stellar
winds, recent analytic formalisms for the mass-loss rates are applied to
the very massive models ($M_{\rm i} \ge 120 M_{\odot}$).  

\end{abstract}

\section{Introduction}

Over the years the existence of a primeval generation of (very) massive 
stars has been invoked in relation to various astrophysical issues
(see Carr et al. 1984; Weiss et al. 2000 for extended reviews), such 
as:
\begin{itemize}
\item fill the gap between the chemical abundances predicted by the
Big Bang nucleosynthesis and those measured in Population-II stars;
\item offer a viable solution to the G-dwarf problem 
in the solar neighbourhood;
\item explain the observed enhancement in $\alpha$-elements
observed in metal-poor stars, as well as 
the chemical abundances in the inter-galactic medium as inferred from
the spectra of high-$z$ damped Ly-$\alpha$ systems;
\item contribute to the cosmological helium abundance;
\item cause the re-ionization of the early Universe; 
\item provide a source of dark matter, in form of stellar remnants, 
as required for galactic haloes and galaxy clusters. 
\end{itemize}

Several models of Population-III stars were constructed in the '80
(e.g. El Eid et al. 1983, Ober et al. 1983).
In recent years, the renewed interest from the cosmologic field 
has again spurred several studies on primordial stellar evolution
(e.g. Marigo et al. 2001; Heger \& Woosley 2002). 

\section{Evolutionary stellar models}

In this context we have carried out extensive evolutionary calculations, 
over a large range of stellar masses  
($0.7 M_{\odot} \la M_{\rm i} \la 1000  M_{\odot}$), covering the H- and
He-burning phases, and allowing for a moderate overshooting from convective
cores ($\Lambda = l/H_{p} = 0.5$). Rotation has not been included. 
Stellar tracks are computed under the assumption of constant-mass evolution.
Additionally, for very massive models, with $M_{\rm i} \ge 120 M_{\odot}$, 
we apply recent mass-loss rate ($\dot M$) prescriptions to account for i) 
the radiation-driven winds at very low metallicities (Kudritzki 2002), 
and ii) the intensification effect on $\dot M$ caused by stellar rotation 
(Maeder \& Meynet 2000).

\section{The critical mass $M_{\rm up}$}

\begin{figure}
\plotfiddle{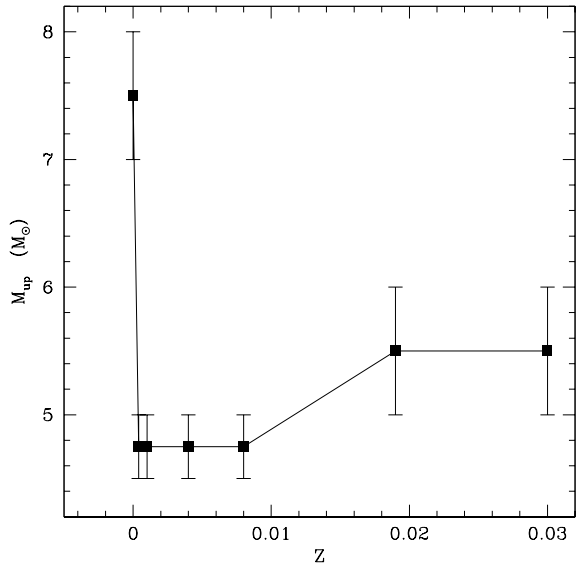}{2.5in}{0}{40}{40}{-140}{-80}
\vskip 0.3cm
\caption{Expected behaviour of $M_{\rm up}$ as a function of metallicity.
The prediction of this work for $Z=0$ is combined 
to the results of Girardi et al. (2000) for other metallicities } 
\label{fig_mup}
\end{figure}

Let us first consider the critical mass $M_{\rm up}$, that is the
usually defined as the maximum initial mass for a star to develop
an electron-degenerate C-O core, hence marking the boundary the limit
between the class of low-intermediate-mass stars and that of massive stars.
We find that it is around $ \sim 7-8 M_{\odot}$ for $Z=0$.

Figure 1 displays the expected trend of  $M_{\rm up}$
as a function of metallicity. It is essentially controlled by the mass 
of the He-core left at the end of the main sequence phase.
At decreasing $Z$ and for CNO-dominated H-burning, 
the initial slight decline is due to the development
of larger convective cores, because of larger luminosities and
more concentrated energy sources.  
Then, the steeper rise at $Z=0$  reflects a reversed
situation, i.e smaller convective cores because of more extended burning
regions. This is caused by the weaker temperature dependence 
energy generation rate when 
the p-p chain becomes competitive with the CNO-cycle, and this latter 
operates at higher temperatures as in the case of $Z=0$ (or extremely low $Z$).

\section{The onset of the triple-$\alpha$ reaction}

\begin{figure}
\plotfiddle{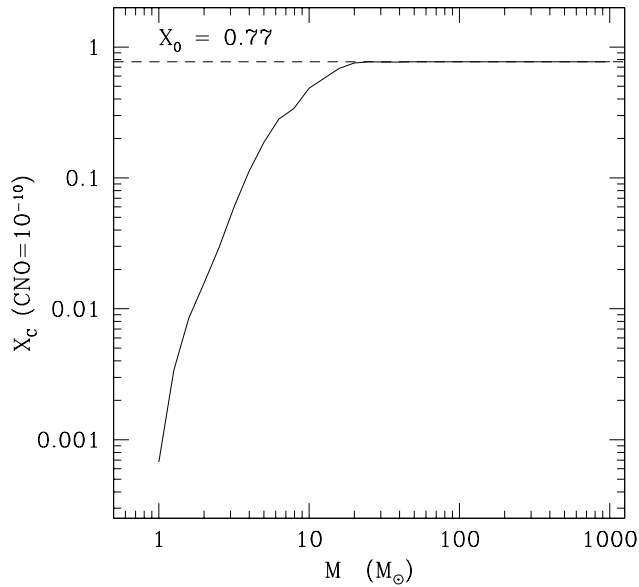}{2.5in}{0}{50}{50}{-140}{-120}
\vskip 1cm
\caption{Evolutionary stage at the onset of the triple-$\alpha$ 
reaction, as a function of the initial mass $M_{\rm i}$.
The ordinate reads the central hydrogen abundance when  
a CNO abundance of $10^{-10}$ (in mass fraction) is built up.} 
\label{fig_3a}
\end{figure}

A peculiarity of $Z=0$ stars is that, 
differently from those with ``normal'' chemical composition, 
the onset of the triple$-\alpha$ reaction
can already happen  during the H-burning phase (see Marigo et al. 2001 for 
a detailed discussion). 
In few words, due to the lack of metals, initially  the CNO-cycle
cannot operate. The only available energy sources are the gravitational
contraction and the p-p chain. Due to the rather weak temperature-dependence
of the latter, the central regions can contract until quite high temperatures
are reached ($\sim 1-2 \times 10^{8}$ K), which lead to the onset of the
$\alpha (2\alpha, \gamma) ^{12}$C reaction.   
As soon as a tiny abundance of $^{12}$C 
is synthesised, then  the CNO-cycle can be activated for the first time.
As shown in Fig.~2, such occurrence depends on the stellar mass,
namely it takes place earlier and earlier at increasing $M_{\rm i}$.
Specifically, in the high-mass domain the triple$-\alpha$ reaction
ignites as soon as the star settles on the zero-age  main sequence,
before burning any significant amount of hydrogen. 

\section{The location in the H-R diagram}

Figure 3 illustrates evolutionary tracks in the H-R diagram for 
selected values of the initial mass, under the assumption of constant-mass
evolution. Basing on these stellar tracks, a set of isochrones 
has been constructed (see Fig.~4). It is available in electronic
fromat at the web-address http://pleiadi.astro.it/.

With respect to massive models  ($M_{\rm i } \ga 8 M_{\odot}$) 
we can make the following remarks:
i) the  H-burning phase is located at quite high effective temperatures 
such that, in combination with high luminosities, we expect these stars to
emit large fluxes of UV photons (see Schaerer 2002);
ii) the He-burning phase takes place at decreasing effective temperatures 
at increasing stellar mass;
iii) stars with $8 M_{\odot} < M_{\rm i} < 70 M_{\odot}$ remain always 
confined in the blue, and are not able to perform any redward excursion 
onto the Hayashi line before the ignition of central carbon.
This will have important consequences for the expected surface chemical changes
(see Sect.~6.).

\begin{figure}
\plotfiddle{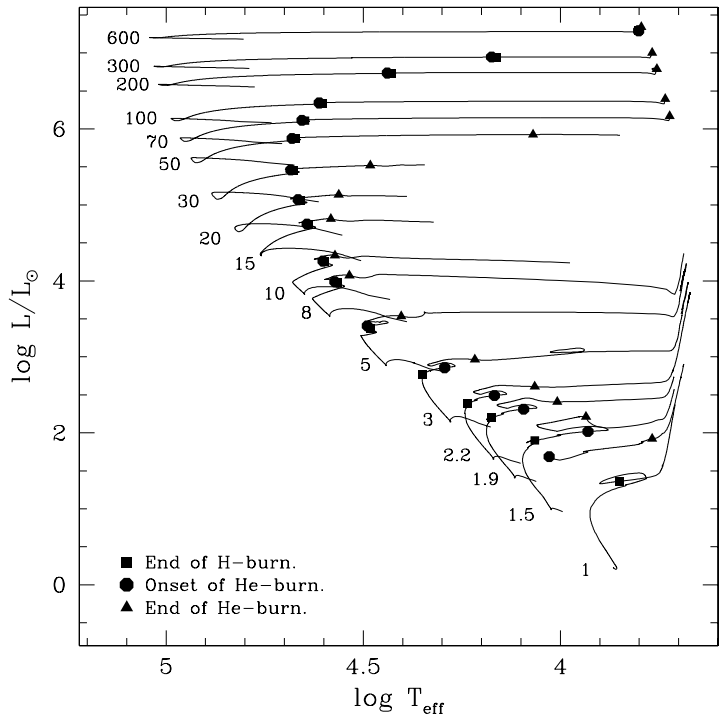}{2.5in}{0}{50}{50}{-160}{-160}
\vskip 2.8cm
\caption{Evolutionary tracks in the H-R diagram for selected values
of the initial stellar mass. Relevant stages are marked along the tracks}
\end{figure}

\begin{figure}
\plotfiddle{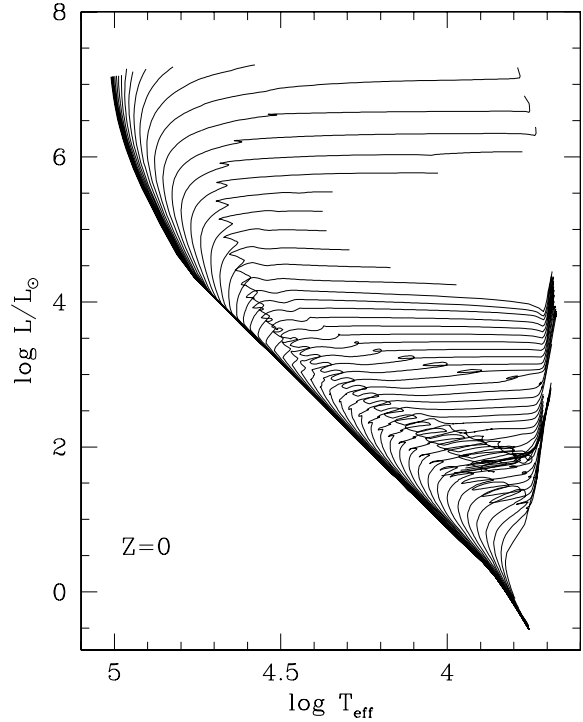}{2.5in}{0}{50}{50}{-160}{-150}
\vskip 2.8cm
\caption{  
Theoretical isochrones in the HR diagram for the initial 
composition $[Z=0, Y=0.23]$. Ages span the range 
from $\log (t/yr)=5.0$ to 10.2, at equally spaced intervals of
$\Delta \log (t/yr) = 0.1$. In all isochrones, the main sequence is
complete down to a stellar mass of 0.7 $M_{\odot}$.}
\end{figure}

\begin{figure}
\plotfiddle{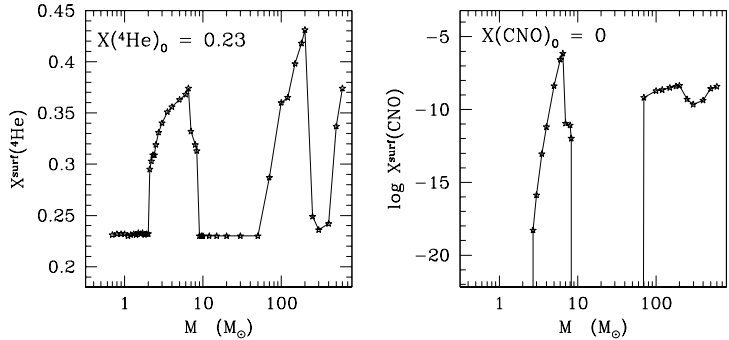}{2.5in}{0}{60}{60}{-180}{-200}
\vskip -2.1cm
\caption{Surface abundances as a function of the stellar mass
in the case of constant-mass evolution. The stage corresponds to the
onset of the first thermal pulses on the AGB for low- and intermediate-mass
stars, and the onset of carbon-burning for massive and very massive stars } 
\label{fig_chim}
\end{figure}

\section{Surface chemical changes}
\label{sec_chim}

As we do not account for any rotationally-induced mixing, changes in the 
surface chemical composition are conditioned to the redward evolution
of the star onto its Hayashi line (see Fig.~3). 
Correspondingly, a deep convective envelope develops and extends into 
a chemically-variable profile (left by the receding convective core during the 
H-burning phase), so that newly-synthesised elements are dredge-up to the 
surface. Predictions for constant-mass evolution are shown 
in Fig.~5

In summary, we find that the $1^{\rm st}$ dredge-up is experienced by stars
with $0.7 M_{\odot} \la M_{\rm i} \la 1.1 M_{\odot}$, whereas the 
$2^{\rm nd}$ dredge-up takes place in stars with 
$2.2 M_{\odot} \la M_{\rm i} \la 8.0 M_{\odot}$, and 
$M_{\rm i} \ga 70 M_{\odot}$. 
It should be noticed that further chemical
changes may occur in the case of mass loss (see Sect.~7.).

\section{Mass loss and other evolutionary properties}

Mass loss is a crucial, but still uncertain, issue 
for primordial stellar evolution.
In this work, we consider two possible driving-processes, namely:
radiative line acceleration and stellar rotation.
To this aim, we adopt analytic formalisms recently presented
by Kudritzki (2002), and Maeder \& Meynet (2000), respectively.

In summary we find that, for a metallicity as low as 
$Z = 10^{-4} \times Z_{\odot}$, radiation is not an efficient
mass-loss mechanism, except for very massive stars with 
$M{\rm i} > 500 M_{\odot}$. 
 
Let us consider Fig. 6, referring to a $250 M_{\odot}$ model
to which we apply  Kudritzki's (2002) prescription.
In this particular case, the amount of mass ejected during the evolution
from the ZAMS to the onset of carbon-burning is negligible, 
i.e. $\sim 0.7 M_{\odot}$. Note the large extension (in mass coordinate) 
of the convective cores during nuclear burnings, and the continuous transition
from the end of core H-burning  to the onset of core He-burning.
In practice, we do not expect any intermediate H-shell burning phase 
in zero-metallicity (very) massive models, due to the 
already high temperatures reached in the central regions at the 
stage of exhaustion od central hydrogen.

Close to the end of the main sequence, the  $250 M_{\odot}$ model evolves
towards lower effective temperatures until it approaches its Hayashi line.
Consequently,  an extended convective envelope develops so that the surface
chemical composition is polluted with nuclear products of H-burning
(mainly $^{4}$He and $^{14}$N).

 \begin{figure}
\plotfiddle{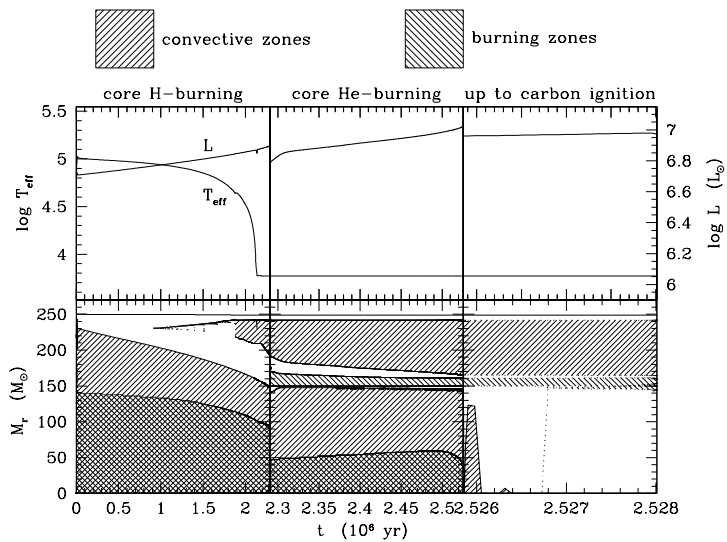}{3.5in}{0}{50}{50}{-150}{-30}
\vskip -2.0cm
\caption{Evolutionary properties of the 250 $M_{\odot}$ model with 
the prescription for the purely radiative mass loss.
Top panel: Stellar luminosity and effective temperature as a function
of time during the major nuclear burnings up to central carbon ignition. 
Bottom panel: Convective and burning regions (in mass coordinate 
from the centre to the surface) across the stellar 
structure. The upper solid line represents the mass coordinate 
of the stellar surface}
\end{figure}
  
Stellar rotation is expected to increase the purely-radiative mass-loss rates.
As already discussed by Meynet \& Maeder (2001), 
the almost total preservation of the initial angular momentum may 
favour the attainment of the break-up velocity  ($\Omega$-limit).
In combination with the very large luminosities of very massive stars, the 
$\Omega\Gamma$-limit might also be approached.
However, it should be noticed that (see  Heger \&
Langer (1998) for an extensive discussion), since a large fraction 
of the total angular momentum is deposited in the outermost layers, 
the ejection of a relatively small amount of mass  from the surface should 
make the star to get rid of a significant part of its angular momentum, 
with consequent spinning down 
of the rotational velocity. As a consequence, the rotational effect on the
mass-loss rate should also weaken. 

The effectiveness of stellar rotation in driving mass loss 
from zero-metallicity stars will be examined in more  detail 
in a forthcoming paper (Marigo et al., in preparation). 
This is an important issue
for very massive stars, in relation to the possible occurrence of 
pair-instability supernovae (Heger A. \& Woosley 2002).

\acknowledgements
P.M. is grateful to N. Langer for providing his expertise on stellar
rotation and helpful remarks on this work.  

\end{document}